\documentstyle[12pt]{article}
\def\dspace{\baselineskip = .30in}

\begin{document}

\title{A SUSY SO(10) Model with Large $tan \beta$}

\author{{\bf G. Lazarides and C. Panagiotakopoulos}\\Physics
Division\\ School of Technology\\University of
Thessaloniki\\Thessaloniki, Greece}

\date{ }
\maketitle

\dspace
\centerline{\bf Abstract}
\vspace{.2in}
We construct a supersymmetric SO(10) model with the asymptotic
relation tan$\beta \simeq m_t/m_b$ automatically arising from its
structure. The model retains the significant  Minimal Supersymmetric
Standard Model predictions for $sin^2 \theta_w$ and $\alpha_s$ and
contains an automatic $Z_2$ matter parity. Proton decay through $d=5$
operators is sufficiently suppressed. It is remarkable that no global
symmetries need to be imposed on the model.

\newpage
\par
The renormalization group  equations of the minimal
supersymmetric standard model (MSSM) with a supersymmetry breaking
scale $M_s \sim 1$ TeV are remarkably consistent$^{(1)}$ with the measured
values of $sin^2 \theta_w$ and $\alpha_s$ and  unification of the three
gauge couplings at a scale $\sim 10^{16}$ GeV. This significant property
certainly makes the MSSM  more
attractive than the non-supersymmetric standard model in spite of the
larger number of undetermined parameters that it contains. An
important undetermined new parameter introduced by supersymmetry, known as $tan
\beta$, is the
ratio of the vacuum expectation values (vevs) of the electroweak higgs
doublets $h^{(1)}$ and $h^{(2)}$ giving mass to the up-type quarks and
the down-type quarks and charged leptons respectively. tan$\beta$
remains undetermined if MSSM is embedded in the minimal supersymmetric
(SUSY) SU(5) model. In contrast, embeddings in supersymmetric grand
unified theories (SUSY GUTs) based on larger gauge groups may lead$^{(2)}$ to
the  asymptotic relation $tan \beta \simeq m_t/m_b$. In a previous
paper$^{(3)}$, we formulated widely applicable conditions under which this
asymptotic relation holds. A central role in these conditions is
played by the $SU(2)_R$ gauge symmetry.
Moreover, we constructed specific SUSY GUTs, mostly based on
semisimple gauge groups,
where these conditions apply.
 The only simple gauge group briefly
discussed was SO(10). However, this discussion did not intend to
produce a specific SO(10) model but
only to indicate the applicability of our conditions.

\par
The purpose of the present paper is to construct a specific SUSY SO(10) model
where the asymptotic relation $tan \beta \simeq m_t/m_b$ is an
automatic consequence of the structure of the theory with the usual
MSSM predictions for $sin^2 \theta_w $ and $\alpha_s$ being retained.
Two problems closely related to the prediction of large tan$\beta$
in this context are
the gauge hierarchy problem and the problem of proton decay
proceeding through
$d=5$ operators$^{(4)}$. Here we do not attempt to provide the mechanism
forcing one pair of electroweak higgs doublets to remain light but we
are satisfied if the relation tan$\beta \simeq m_t/m_b$ holds
 independently of the reason for which one such pair
remains light. As far as
the proton decay problem is concerned, we simply aim at loosening the
usual tight relationship of minimal SUSY GUTs between colour triplet
higgsino mediated proton decay amplitude and light fermion masses.
Because of this relationship
the proton decay
amplitude mediated by colour triplet higgsino exchange is marginally
compatible with the present experimental bound on proton lifetime.
Moreover, large values of $tan \beta$ implied by the asymptotic
relation $tan \beta \simeq m_t/m_b$ make the situation even  worse.
Any viable SUSY SO(10) model with large $tan\beta$ must therefore be
equipped
with a mechanism which suppresses the $d=5$ operators relevant for
proton decay mediated by colour triplet higgsino exchange. The usual
mechanism$^{(5)}$ for achieving such a suppression involves
arranging for the colour triplets which belong to
the same SO(10) multiplets with the electroweak higgs doublets to
acquire  masses close to the unification scale but with mass partners
having suppressed couplings to the ordinary quarks and leptons. This
arrangement, in most cases, requires non-minimal field content,
additional discrete or continuous global symmetries and, sometimes,
even a mild tuning of parameters. For our purposes significant
departure from
minimality is certainly undesirable because it reduces the
predictability of the model. However, we succeed here in implementing
the above described mechanism
with a minimal enlargement of the field content of the
theory, with
only a mild tuning of  parameters (by an order of
magnitude or so)
and with the
asymptotic relation $tan \beta \simeq m_t/m_b$ being  automatically
guaranteed.
We find particularly worth emphasizing  the fact that we do not
have to recourse to the imposition  of any discrete symmetries since
all the above  achievements are immediate consequences of the gauge
symmetry SO(10) and the choice of the field content.

\par
It is quite remarkable that even the discrete matter parity, necessary
for suppressing the rapid proton decay through $d=4$ operators, could
be an automatic consequence of the structure of the theory. Any SO(10)
model has an obvious $Z_2$ symmetry under which the spinors change
sign with the tensors remaining invariant. If no spinor acquires
a vev this $Z_2$ symmetry remains unbroken and plays the role
of the matter parity provided the three light generations belong to
three 16's of SO(10). It is easily seen that the $Z_2$ matter parity
just described is a subgroup of both $U(1)_{B-L}$ and of the $Z_4$
center of SO(10).

\par
We consider a SUSY SO(10) model with the ordinary quarks and leptons
belonging as usual to three 16's denoted by $\psi$. The SO(10) breaks
directly down to $SU(3)_c \times SU(2)_L \times U(1)_Y$. This is
achieved by using one superfield in each of the  126, $\overline
{126}$, 45 and 54  representations of SO(10). We  denote these fields
as $\chi_{ijklm}, \bar {\chi}_{ijklm}, \phi_{ij}$ and $\eta_{ij}$
respectively. Note that $\chi_{ijklm} (\bar{\chi}_{ijklm})$ is a fully
antisymmetric and (anti)self-dual SO(10) tensor whereas
$\phi_{ij}(\eta_{ij})$ is an antisymmetric (symmetric and traceless) SO(10)
tensor. The superpotential terms allowed by the SO(10) symmetry are
$\bar{\chi}_{ijklm} \chi_{ijklm}, \phi_{ij} \bar{\chi}_{iklmn}
\chi_{jklmn}$,
$\eta_{ij} \chi_{iklmn} \chi_{jklmn},\eta_{ij}
\bar{\chi}_{iklmn} \bar{\chi}_{jklmn}$,
$ \phi_{ij} \phi_{ij},
\eta_{ij} \eta_{ij}, \eta_{ij} \eta_{jk}\eta_{ki} $ and $\eta_{ij} \phi_{jk}
\phi_{ki}$.
It can be shown$^{(6)}$ that, with these fields, the SO(10) breaking proceeds
without leaving any accidental pseudogoldstone particles. Thus, all
particles in this sector of the theory can be given masses close to
the unification scale. This fact is important for retaining the
successful MSSM predictions for $sin^2 \theta_w$ and $\alpha_s$.

\par
In order to achieve the  electroweak breaking in a way that allows us to
implement the above mentioned mechanism for suppressing proton decay
through $d=5$ operators, we introduce two more superfields, one in the
10 and one in the $210^\prime$  representation of SO(10). These
fields are  denoted as $\zeta_i$ and $ \theta_{ijk}$ respectively.
$\zeta_i$ is a SO(10) vector and
$\theta_{ijk}$ is a fully symmetric and traceless SO(10) tensor.
Both $\zeta$ and $\theta$ contain components which
have the right standard model quantum numbers to contribute to the
electroweak doublets but only $\zeta$ has yukawa couplings to the $\psi$'s
which contain the ordinary quarks and leptons. The fields $\zeta$ and
$\theta$ also contain colour triplets and antitriplets. So, if one
succeeds in arranging for the  colour triplets and antitriplets in
$\zeta$ to
have as mass partners
mainly the colour  antitriplets and triplets  in $\theta$
respectively,
one can suppress the $d=5$ operators
for proton decay. The relevant mass term is provided by the
superpotential term $ \lambda \zeta_i \theta_{ijk} \eta_{jk}$ and must be
the dominant mass term between these colour triplets and antitriplets
for this
mechanism to work. The other mass terms between these triplets and
antitriplets
provided by the superpotential terms $M \zeta_i \zeta_i, \lambda^\prime
\eta_{ij} \zeta_i \zeta_j, M^\prime \theta_{ijk} \theta_{ijk}$ and
$\lambda^{\prime \prime}
\eta_{ij} \theta_{ikl} \theta_{jkl}$ must be slightly suppressed (by
an order of magnitude or so). An explicit calculation of the mass
matrix between the colour triplets and antitriplets in $\zeta$ and $\theta$
which belong to the (6,1,1) representation of $ G_{422} \equiv SU(4)_c \times
SU(2)_L
\times SU(2)_R$ gives $M + \lambda^\prime \eta_{o}/6$, $ M^\prime -
\lambda^{\prime  \prime} \eta_{o}/54$
in the main diagonal and $\lambda \eta_{o}/3
\sqrt{2}$ in the off-diagonal entries. Here $\eta_o$ is defined by the
vev of $\eta_{ij} : <\eta_{ij}> = \eta_o$ diag (1/6,...,1/6, -
1/4,...,-1/4), where the  1/6 appears six times and the  -1/4 four
times. The mass matrix of the electroweak doublets in $\zeta$ and $\theta$,
which happen to belong to the (1,2,2) representation of
$G_{422}$,  consists of $ M - \lambda^\prime
\eta_{o}/4$, $M^\prime - \lambda^{\prime \prime} \eta_{o}/9$ in the
main  diagonal and $-\lambda \eta_{o}/4$ in the off-diagonal entries.
One can also compute the masses of the rest of the components of
$\theta$. They turn out to be $M^\prime + \lambda^{\prime \prime}
\eta_{o}/6$, $ M^\prime - \lambda \eta_{o}/9$, $M^\prime -\lambda
\eta_{o}/4$ and $M^\prime + \lambda \eta_{o}/36$ for the (50,1,1),
(6,3,3), (1,4,4) and $(20^\prime$,2,2) components respectively. Here
again
the decomposition of $\theta$ is under $G_{422}$.
The electroweak
pair of doublets is chosen to lie in an arbitrary generic linear combination
of the doublets in $\zeta$ and $\theta$ by fine tuning one eigenvalue of the
corresponding $2 \times 2$ mass matrix. It is then obvious that the
diagonal colour (anti)triplet mass terms can be slightly suppressed
without affecting any physical masses in the theory. This fact
combined with the statement at the end of the previous paragraph means
that all the states in the theory except the MSSM states (and the
three right-handed neutrinos) can acquire masses close to the
unification scale. Consequently, the significant MSSM prediction for
$sin^2 \theta_w$ and $\alpha_s$ is retained with the $d=5$ operators
for proton decay being sufficiently suppressed at the same time.

\par
It is important to note that the fields $ \bar{\chi}, \chi$ also
contain candidate electroweak doublets. Always restricting ourselves to
renormalizable superpotential terms we find that these doublets do
not couple to the doublets in $\zeta$ and $\theta$ by mass terms.This
is due to the absence of any trilinear superpotential terms coupling
$\zeta$ or $\theta$ with $\bar{\chi}$ or $\chi$ and the fields
$ \bar{\chi},
\chi, \phi $ or $\eta$ acquiring large vevs.
So the doublet sector in $\zeta$ and $\theta$
does not mix, through renormalizable interactions, with the doublets
in $\bar{\chi}, \chi$. Note that, in principle, we had the option to
choose the MSSM pair of higgs doublets from the $\bar{\chi}, \chi$ sector
rather than from the $\zeta, \theta$ sector. We avoided this
possibility since we prefer to keep the successful asymptotic relation
$m_b \simeq m_\tau$ over the relation $m_b \simeq \frac {1}{3} m_\tau$.
Moreover,
the mechanism we employed for suppressing
proton decay through $d=5$ operators would be of no use with this choice.

\par
The mass matrix  of the electroweak doublets in $\zeta$ and $\theta$
has an important property: it does not get renormalizable contributions
from vevs which break the $SU(2)_R$ (and the $SU(4)_c$)
 symmetry. Such vevs are acquired
by $\bar{\chi}, \chi$ and $\phi$ but trilinear couplings of $\zeta$ or
$\theta$ with $\zeta$ or $\theta$ and one of the $\bar{\chi}, \chi $ or
$\phi$ are not allowed by the SO(10) symmetry. This means that the unique
pair of electroweak doublets forms a single $SU(2)_R$ -doublet.
Corrections from non-renormalizable terms cannot alter this
significant property since the orthogonal pair of heavy doublets,
which are also $SU(2)_R$ -partners, is not affected by such small
contributions.

\par
Since  the three ordinary light generations belong  entirely to the
three $\psi$'s, it is obvious that the  $SU(2)_L$ - doublet quarks and
leptons are $SU(2)_R$ - singlets with the $SU(2)_L$ - singlet
antiquarks and antileptons being $SU(2)_R$ - doublets. This
observation combined with the facts that the unique electroweak higgs
doublet pair $h^{(1)}, h^{(2)}$ forms (to a very good approximation) a
$SU(2)_R$ -doublet and that the  quark and lepton tree-level (Dirac)
mass terms must be (to the same approximation) $SU(3)_c \times SU(2)_L
\times SU(2)_R \times U(1)_{B-L}$ (and even $G_{422})$
invariant implies that $tan\beta \simeq  m_t/m_b
\simeq m^D_{\nu_{\tau}}/m_\tau$ (and $m_b \simeq
m_\tau $).
Analogous
relations for the first two families are not expected to hold since we
anticipate for them substantial corrections to the tree-level masses.

\par
One readily reaches the same conclusion with immediate application of
the proposition stated in ref.(3).
{}From the above discussion is obvious that the only states which remain
massless in the supersymmetric limit after the breaking of SO(10) are
the MSSM states with all other states acquiring superheavy masses.
Also, the unique pair of electroweak doublets forms a
$SU(2)_R$ -doublet and the ordinary quarks and leptons are unique under
(matter parity) $\times$ (standard model group). Besides the $Z_2$
matter parity belongs to the center of SO(10).
The conditions stated in
ref.(3) are then satisfied  and, consequently, the approximate
relation $tan \beta \simeq m_t/m_b$ holds. In addition, there are
exactly three standard model singlet states which can play to role of
the right-handed neutrinos. They are the standard model singlets in
the $\psi$ 's. We then also have $tan \beta \simeq
m^D_{\nu_{\tau}}/m_\tau$
with $m^D_{\nu_{\tau}}$ being the Dirac mass of the $\tau$-neutrino.

\par
The fields $\bar{\chi}$ and $\chi$ also contain colour triplet and
antitriplet components which form a (6,1,1) representation of
$G_{422}$. Due to the couplings $ f \psi \psi
\bar{\chi}$ (and the couplings $f^\prime \bar{\chi} \bar{\chi} \eta,
f^{\prime \prime} \chi \chi \eta$)
the triplets and antitriplets in $\bar{\chi}$
give rise to $d=5$ operators for
proton decay. The same yukawa couplings are also responsible for the
Majorana masses of the right-handed neutrinos  and are certainly less
constrained than the yukawa couplings $ \psi \psi \zeta$ which
contribute
to ordinary quark and lepton masses. Values of the $f$'s on the order
of $10^{-4} -10^{-5}$ are sufficiently small for suppressing proton
decay through $d=5$ operators and for obtaining  right-handed neutrino
masses of order $10^{11} -10^{12}$ GeV. The $\tau$-neutrino can then
be cosmologically significant. Additional suppression of these $d=5$
operators could be obtained by some suppression of the $f^\prime$ and
$f^{\prime \prime}$ couplings.

\par
The problem of proton decay through $d=5$ operators mediated by
exchange of colour triplets and antitriplets in $\bar{\chi}$ could be
more elegantly addressed be employing discrete symmetries. Let us
introduce a $Z_4$ discrete symmetry the generator of which acts on
 $\chi$ and
$\bar{\chi}$  as i and -i respectively and leaves all other fields
invariant. This symmetry forbids terms like $\bar{\chi}
\bar{\chi} \eta$ and $\chi \chi \eta$ but unfortunately also the terms
$\psi \psi \bar{\chi}$ responsible for right-handed neutrino Majorana
masses. To deal with this problem, we further introduce a pair of
SO(10) singlets $S, \bar{S}$ on which
the $Z_4$ symmetry generator acts also as i and -i
 and which acquire a superheavy vev as
well. Now the terms $\psi \psi \bar{\chi} <S> /M_P$ (where $M_P \sim
10^{19}$ GeV is the Planck mass) are more than sufficient for right-
handed neutrino Majorana masses, whereas the combined effect of these
terms and the more suppressed terms $\bar{\chi} \bar{\chi} \eta
<S>^2/M^2_P, \chi \chi \eta <S>^2/M^2_P$ give rise to sufficiently
suppressed proton decay rate through d=5 operators involving exchange
of the colour triplets and antitriplets in $\bar{\chi}$.

\par
By employing  discrete symmetries one can construct SO(10)
models with tan$\beta \simeq m_t/m_b$ in which the $\chi, \bar{\chi}$
fields transforming as 126, $\overline{126}$ under SO(10) are replaced
by spinors $\psi^\prime, \bar{\psi}^\prime$ transforming as 16,
$\overline{16}$. In this case, first of all, the automatic $Z_2$
matter parity is broken and another matter parity has to be introduced
as an additional discrete symmetry. This is easily done by assigning
minus matter parity to the three $\psi$'s and plus matter parity to
all other fields. The mechanism of suppression of d=5 operators by the
introduction of the $210^\prime$ representation goes through here as
well. Moreover, we do not have to worry about any other $d=5$
operators apart from the ones
due to the exchange of colour triplets and antitriplets in
$\zeta$. The only additional problem
destroying the large tan$\beta$ prediction is that there is a
potential mixing between the electroweak doublets in the vector
$\zeta$ (which are $SU(2)_R$ - doublets) and the electroweak
doublets in the spinor $\psi^\prime$ (which are $SU(2)_R$-singlets)
through the $SU(2)_R$ -breaking vev $<\psi^ \prime>$, the relevant term
being $\zeta \psi^ \prime <\psi^\prime>$. To forbid this term we
introduce a $Z_3$ symmetry under  which the only
non-invarinant fields are $\psi^\prime, \bar{\psi}^\prime$ and a gauge
singlet $S$. These fields transform under the generator of $Z_3$
by multiplication with
$\alpha, \alpha^2$ and $\alpha^2$ respectively ($\alpha=e^{2 \pi
i/3}$). The singlet $S$ acquires a vev whose natural value is $\sim
M^{1/3}_P
M^{2/3} (M \sim 10^{16}$ GeV being the only superheavy mass scale in
the model). Then the terms $\psi \psi \bar{\psi}^\prime
\bar{\psi}^\prime <S>/M_P$ give
acceptable Majorana masses to the right-handed neutrinos
while the term $\zeta \psi^\prime \psi^\prime
<S>^2/M^2_P$ generates a sufficiently small $SU(2)_R$ -breaking mixing between
electroweak doublets in $\zeta$ and $\psi^\prime$. Therefore, again,
the light electroweak higgs doublets, which must necessarily be chosen
from a linear combination of $\zeta$ and $\theta$ (since $\psi^\prime$
does not couple to quarks and leptons), form (to a good approximation) a
$SU(2)_R$ -doublet and our relation for tan$\beta$ holds.

\par
We presented a SO(10) SUSY GUT with the asymptotic relation $tan \beta
\simeq m_t/m_b$ automatically arising from the structure of the
theory. A $Z_2$ matter parity also emerges automatically and the
significant MSSM predictions for $sin^2 \theta_w$ and $\alpha_s$ are
retained. The $d=5$ operators for proton decay are sufficiently
suppressed. All this is achieved without imposing any global
symmetries and with a relatively simple field content.
Alternatives employing additional discrete symmetries were also
briefly discussed.

\newpage
\section*{References}

\begin{enumerate}

\item S. Dimopoulos and H. Georgi, Nucl. Phys. \underline{B193} (1981)
150;
J.Ellis, S.Kelley and D.V.Nanopoulos, Phys. Lett.
\underline{B249} (1990) 441; U. Amaldi, W.de Boer and H. Furstenan,
Phys. Lett. \underline{B260} (1991) 447; P.Langacker and M.X.Luo,
Phys. Rev. \underline{D44} (1991) 817.

\item B. Ananthanarayan, G. Lazarides and Q. Shafi, Phys. Rev.
\underline{D44} (1991) 1613; H.Arason, D.J.Casta\~{n}o, B.E. Keszthelyi,
S. Mikaelian, E.J.Piard, P.Ramond and B.D. Wright, Phys. Rev. Lett.
\underline{67} (1991) 2933; S. Kelley, J.L.Lopez and D.V.Nanopoulos,
Phys. Lett. \underline{B272} (1992) 387.

\item  G. Lazarides and C. Panagiotakopoulos, Thessaloniki
Univ.preprint UT-STPD-1-94(1994).

\item N.Sakai and T.Yanagida, Nucl.Phys.\underline {B197} (1982) 533; S.
Weinberg, Phys. Rev. \underline{D26} (1982) 287; J. Hisano, H.
Marayama and T. Yanagida, Nucl.Phys. \underline{B402} (1993) 46.

\item J. Hisano, H. Murayama and T. Yanagida, Phys. Lett. \underline
{B291} (1992) 263;  K.S. Babu and S.M.Barr, Phys.Rev. \underline {D48} (1993)
5354.

\item D.G.Lee and R.N. Mohapatra, private communication.
\end{enumerate}

\end{document}